\input{aipcheck}

\documentclass[
    final            
  ]
  {aipproc}

\usepackage{footmisc}
\usepackage{floatflt}
\input babarsym
\layoutstyle{6x9}

\begin{document}
\title{Searches for ${D}^{0}$-$\bar{D^{0}}$ Mixing, Rare Charm and Tau Decays\footnote{~to
appear in Proceedings of SUSY06, the 14th International Conference on Supersymmetry and the 
Unification of Fundamental Interactions, UC Irvine, California, 12-17 June 2006.}}


\author{Sanjay K. Swain}{
  address={2575 Sand Hill Road, Stanford Linear Accelerator Center, Menlo park, CA-94025}
}

\begin{abstract}
 I discuss the results on $D^{0}$-$\bar{D}^{0}$ mixing through hadronic as well as semi-leptonic 
charm decays, rare flavor-changing neutral currents in the charm sector and the lepton flavor 
violating $\tau$ decaying to charged lighter leptons. The results from both \babar\ and Belle 
are presented in this review.
\end{abstract}

\maketitle


\section{$D^{0}$-$\bar{D}^{0}$ mixing via $D^{0} \to K^+\pi^-\pi^{0}$, $K^+\pi^-$ and $K^{(*)-}e^{+}\nu$}
The mixing in the charm sector, namely $D^{0}-\bar{D}^{0}$ mixing has 
received less attention in the past because of very small Standard Model (SM) expectations.
However, it is the GIM (Glashow, Iliopoulos and Maiani) mechanism that makes the charm 
mixing so interesting. At small distances, this mixing proceeds via flavor-changing
neutral currents (FCNC). Since there are no tree-level FCNC contributions
in the SM, processes such as mixing occur at the quark level
primarily via box diagrams. The exact evaluation of box diagram is not so important
because $D^{0}-\bar{D}^{0}$ mixing is probably dominated by long distance effects~\cite{DMixTheory}, 
i.e., by intermadiate hadronic states (not quarks) in the $D^{0}-\bar{D}^{0}$ transition.

The mixing processes are parameterized with the quantities $x$ and $y$ where 
\begin{equation}
\label{eq:xydef}
x \equiv 2\frac{m_{2} - m_{1}}{\Gamma_{2} + \Gamma_{1}},\ \ \ \ \ %
y \equiv \frac{\Gamma_{2} - \Gamma_{1}}{\Gamma_{2} + \Gamma_{1}}%
\textrm{.}
\end{equation}
$(m_{1},m_{2})$ and $(\Gamma_{1},\Gamma_{2})$ are the mass and decay widths
of mass eigenstates $|D_{1,2}\rangle = p|D^0 \rangle \pm q|\bar{D^0}\rangle$ ($p$ and $q$ are complex numbers). 
If $x$ and $|y|$ are very small, then $D^0$ and $\bar{D^0}$ practically do not oscillate into and from 
each other while decaying; the linear superposition at the production time is identical with the one at the decay time.

The first search for $D$ mixing in the decay $D^{0} \to K^+\pi^-\pi^{0}$, which is wrong sign ($WS$) decay, is
presented here. For a nonleptonic multibody WS decay, the time dependent decay rate, $\Gamma_{WS}(t)$, 
relative to a corresponding right-sign ($RS$) rate, $\Gamma_{RS}(t)$, is
approximated by~\cite{Blaylock:1995ay}
\begin{eqnarray}
 & {\displaystyle\frac{\Gamma_{\small{WS}}(t)}{\Gamma_{\small{RS}}(t)} =%
   \tilde{R}_D + \alpha\tilde{y}'\sqrt{\tilde{R}_D}\,(\Gamma t)%
   + \frac{\tilde{x}'^2+\tilde{y}'^2}{4}(\Gamma t)^2 } & \\
 &  0 \leq \alpha \leq 1\textrm{.} & \nonumber
\end{eqnarray}
The tilde indicates quantities that have been integrated over any choice
of phase-space regions.  $\tilde{R}_D$ is the integrated doubly Cabibbo suppressed (DCS) branching ratio,
$\tilde{y}' = y\cos\tilde{\delta} - x\sin\tilde{\delta}$ and
$\tilde{x}' = x\cos\tilde{\delta} + y\sin\tilde{\delta}$,
where $\tilde{\delta}$ is an integrated strong-phase difference
between the Cabibbo favored (CF) and the DCS decay amplitudes,
$\alpha$ is a suppression factor that accounts for
strong-phase variation over the phase space regions, and $\Gamma$ is the
average width.  The time-integrated mixing rate
$R_M = (\tilde{x}'^2+\tilde{y}'^2)/2 = (x^2+y^2)/2$ is independent of decay mode.
The \CP-violating effects are studied by fitting to the $D^{0} \to K^+\pi^-\pi^{0}$
and $\bar{D^{0}} \to K^-\pi^+\pi^{0}$ samples separately. The $CP$-violation in the
interference between the DCS channel and mixing, parameterized by an integrated 
\CP-violating-phase difference $\tilde{\phi}$, as well as \CP\ violation
in mixing, parameterized by $|p/q|$. The $CP$ invariance in the DCS and CF decay rates is assumed.

Assuming \CP\ conservation, using 230.4 \invfb\ \babar\ measured the time-integrated mixing rate
$R_M =$ (0.023 $\mbox{}^{\rm +0.018}_{\rm -0.014}$\,(stat.) $\pm$ 0.004\,(syst.))\%,
and $R_M <$ 0.054\% at the 95\% confidence level~\cite{Aubert:2006wilson}.
The data is consistent with no mixing at the 4.5\% confidence level (C.L).
Considering the entire allowed phase space, the branching ratio for $WS$ decay
relative to $RS$ decay is (0.214 $\pm$ 0.008\,(stat.) $\pm$ 0.008\,(syst.))\%.

Belle searched for $\Dz$-$\Dzb$ mixing in $WS$ decay $D^{0} \to K^+\pi^-$
 based on 400 \invfb of data~\cite{Zhang:2006dp}. Assuming \CP\ conservation,
Belle finds ${x'^2}<0.72\times10^{-3}$ and $-9.9\times10^{-3}<y'<6.8\times10^{-3}$ at the 95\% C.L.
The no-mixing point ($0,0$) has a confidence level of 3.9\%. Assuming no mixing, $R_D=(0.377\pm0.008\pm 0.005)\%$.
Also Belle searched for mixing in the neutral $D$ meson system using semileptonic $D^0\to K^{(*)-}e^+\nu$ decays using
253 \invfb of data~\cite{Abe:2005nq}. Neutral $D$ mesons from $D^{\ast+}\to D^0\pi^+$ decays are used; the 
flavor at production is tagged by the charge of the slow pion. From the yield of $RS$ and $WS$ decays arising from
non-mixed and mixed events, respectively, the upper limit (UL)of
the time-integrated mixing rate is $R_M < 1.0\times 10^{-3}$ at $90\%$ C.L.

\section{Measurement of the Pseudoscalar Decay Constant \fds Using Charm-Tagged Events}
Measurements of pure leptonic decays of charmed pseudoscalar
mesons are of particular theoretical importance. They provide an unambiguous
determination of the overlap of the wavefunctions of the heavy and light
quarks within the meson, represented by a single decay constant ($f_M$)
for each meson species ($M$). The partial width for a \Ds meson to
decay to a single lepton flavor ($l$) and its accompanying
neutrino ($\nu_l$), is given by
\begin{equation}
  \!\hspace{.3ex}\Gamma(\Ds\to l^+\nu_l)
  = \frac{G^2_F|V_{cs}|^2}{8\pi}  \fds^2m_l^2m^{}_{D_s}\Bigg(1-\frac{m^2_l}{m^2_{D_s}}\Bigg)^2,
  \label{eq:br}
\end{equation}
where $m^{}_{D_s}$ and $m^{}_l$ are the $\Ds$ and lepton masses,
respectively, $G_F$ is the Fermi constant, and $V_{cs}$ is the CKM
matrix element giving the coupling of the weak charged current to the $c$ and $s$ quarks.
In order to measure $\Ds\to\mu^+\nu_\mu$, the decay chain $\Dss\to\gamma\Ds,
\Ds\to\mu^+\nu_\mu$ is reconstructed from \Dss mesons produced in the
hard fragmentation of continuum \ccbar events.
The branching fraction of \dstomunu cannot be determined directly,
since the production rate of $D_s^{(*)+}$ mesons in \ccbar
fragmentation is unknown.  Instead the partial width ratio
$\Gamma(\dstomunu)/\Gamma(\dstophipi)$ is measured by reconstructing
\dsstodstophipi decays. The \dstomunu branching fraction is evaluated
using the measured branching fraction for \dstophipi.
Using the \babar\ average for the branching ratio
$\BR(\dstophipi)=(4.71\pm0.46)\,\%\,$~\cite{Aubert:2005xu}~\cite{Aubert:2006nm},
we obtain the branching fraction $\BR(\dstomunu) = (6.74 \pm 0.83 \pm 0.26 \pm 0.66)\times10^{-3}$
and the decay constant $\fds = (283 \pm 17 \pm 7 \pm 14){\mev}$.
The first and second errors are statistical and systematic,
respectively; the third is the uncertainty from $\BR(\dstophipi)$. 
Using $\BR(\dstophipi)_\mathrm{PDG} = (3.6\pm0.9){\%}$~\cite{Eidelman:2004wy}, the branching fraction 
is $\BR(\dstomunu) = (5.15 \pm 0.63 \pm 0.20 \pm 1.29)\times10^{-3}$ and 
the decay constant $\fds = (248 \pm 15 \pm 6 \pm 31){\mev}$~\cite{Aubert:2006stelzer}.
                                               
\section{Search for Flavor-Changing Neutral-Current Charm Decays}
As mentioned earlier, in SM, FCNC processes cannot occur at the tree level. It therefore provides
an excellent tool for investigating the quantum corrections in the SM
as a way to search for evidence of physics beyond the SM. FCNC
processes have been studied extensively for $K$ and $B$ mesons in
$\Kz-\Kzb$ and $\Bz-\Bzb$ mixing processes and in rare FCNC decays,
such as $s\to d\ellell$, $b\to s\gamma$ and $b\to s\ellell$ decays.
The present measurements of these processes agree with SM predictions~\cite{thurth},
but there are strong ongoing efforts to improve both the measurements
and the theoretical predictions, and to measure new effects, such as
CP violation, in FCNC processes. In the SM very small signals are expected, as a consequence of
effective GIM cancellation. This contribution is masked by the presence of
long-distance contributions from intermediate vector resonances.
\babar\ searched for rare FCNC charm decays of the form $X_c^+\to h^+\llp$,
where $X_c^+$ is a charm hadron, $h$ is a pion, kaon or proton, and
$\ell^{(}{'}^{)}$ is an electron or a muon. In the pion and kaon
modes, both $D^+$ and $D_s^+$ decays are studied, while in the proton
modes $\Lambda_c^+$ decays are studied.  Based on a data sample of
288\invfb, the ULs on the branching fractions of the different decay modes  
 at 90\% C.L are set in the range $(4-40) \times10^{-6}$~\cite{Aubert:2006petersen}.

\section{Lepton Flavor Violation in Tau decays}
The observation of the neutrino oscillation implies that the lepton flavor is not individually 
conserved. Therefore lepton-flavor-violating (LFV) processes
in the charged-lepton sector are allowed. In simple extentsions to SM, which accomodate
mixing of massive neutrinos, the rate of the LFV processes is accompanied
by large suppression and is far out of the reach of the experimental detection.
On the other hand, several SUSY models~\cite{Hisano:1998fj} predict LVF tau decay rates up to and above
the present experimental limits, making that physics channel one of
the most sensitive probes for low-energy SUSY searches.

The results for lepton flavor violating decays $\tau \to l x$ where $l$ is either electron or muon, and 
$x$ is $\gamma$, $K^{0}_{S}$, $hh$ ($h$ is light charged hadrons: kaon or pion ) or $V^{0}$ (a neutral vector mesons 
such as $\rho^0$, $K^*(892)^0$, $\phi$) are discussed in this section.  
\babar\ and Belle searched for the above mentioned LFV tau decays in
tau pairs produced from e+e- collisions around the Y(4s) peak.

In the typical analysis, the plane perpendicular to the thrust axis
is used to divide each event in two hemispheres containing a candidate
SM tau decay (tag-side) and a LFV tau decay (signal-side),
respectively. Background events are suppressed with requirements on
the missing mass, momentum or energy of the event, due to the
neutrino(s) in the SM tau decay. Signal events are selected by
requiring that the candidate decay products total evergy is
compatible with the tau energy in the center-of-mass (C.M.) system (half
the total C.M. energy) and that their invariant mass is compatible with the tau
mass within the experimental resolution.

Using data corresponding to an integrated luminosity of  232 \invfb\, \babar\ measures the UL on branching fraction, 
${\cal{B}}(\tau^-\rightarrow \mu^-\gamma) < 6.8\times 10^{-8}$~\cite{Aubert:2005ye} 
and ${\cal{B}}(\tau^-\rightarrow e^-\gamma) < 1.1\times 10^{-7}$~\cite{Aubert:2005wa}
at the 90\% C.L. Using the results from
$\tau^-\rightarrow \mu^-\gamma$ one can constrain the parameter space in $\tan{\beta}-m_{A}$ plane, where
$\tan\beta$ is the ratio of vacuum expectation values of the two Higgs doublets and the $m_{A}$ is the
\CP\ conserving Higgs boson mass, and the current result excludes the higher $\tan{\beta}$ region.

Belle searched for the lepton flavor violating decays $\tau^-\rightarrow l^-K^{0}_{S}$ ($l= \mu,e$) using 281 \invfb\ data 
and set the UL to be less than $4.9\times 10^{-8}$ and $5.6\times 10^{-8}$ at the 90\% C.L, for
muon and electron modes, respectively~\cite{Miyazaki:2006sx}. For $\tau^-\rightarrow lhh$ or
 $\tau^-\rightarrow lV^{0}$ modes~\cite{Yusa:2006qq}, with 158 \invfb\ data, 
Belle set 90\% C.L upper limits for the branching fractions in the range of $(1.6-8.0) \times 10^{-7}$ at the 90\% C.L. 
For $\tau^-\rightarrow lhh$ decay, using 221 \invfb data, \babar\ set upper limits in the range of $(0.7-4.8)\times 10^{-7}$ 
at 90\% C.L~\cite{Aubert:2005tp}.

\begin{theacknowledgments}
This work is supported Department of Energy (US). I thank 
experts from \babar\ and Belle collaboration for the 
useful discussion and providing the results used in this review.
\end{theacknowledgments}


\bibliographystyle{aipprocl} 


\end{document}